\documentclass[prb,twocolumn,superscriptaddress,showpacs,nofootinbib]{revtex4}
\usepackage{graphicx}% Include figure files

\usepackage{bm}% bold math
\usepackage{amssymb}
\usepackage{epstopdf}
\usepackage{amsmath}
\usepackage{amstext}
\usepackage{latexsym}
\usepackage{hyperref}
\usepackage{pstricks}
\usepackage{ulem}

\newcommand{\ket}[1]{\left\vert#1\right\rangle}
\newcommand{\bra}[1]{\left\langle#1\right\vert}

\begin{document}

\title{The bilinear-biquadratic spin-1 chain undergoing quadratic Zeeman effect}
\author{G. De Chiara}
\affiliation{F\'isica Te\`orica: Informaci\'o i Processos Qu\`antics, Universitat Aut\`{o}noma de Barcelona, E-08193 Bellaterra, Spain}
\affiliation{Kavli Institute for Theoretical Physics, University
of California, Santa Barbara, California  93106-4030}
\author{M. Lewenstein}
\affiliation{ICFO--Institut de Ci\`encies Fot\`oniques,
Parc Mediterrani de la Tecnologia, 08860 Castelldefels, Spain}
\affiliation{ICREA, Instituci\`o Catalana de Recerca i Estudis Avan\c{c}ats, E08011 Barcelona}
\affiliation{Kavli Institute for Theoretical Physics, University
of California, Santa Barbara, California  93106-4030}
\author{A. Sanpera}
\affiliation{F\'isica Te\`orica: Informaci\'o i Processos Qu\`antics, Universitat Aut\`{o}noma de Barcelona, E-08193 Bellaterra, Spain}
\affiliation{ICREA, Instituci\`o Catalana de Recerca i Estudis Avan\c{c}ats, E08011 Barcelona}
\affiliation{Kavli Institute for Theoretical Physics, University
of California, Santa Barbara, California  93106-4030}

\begin{abstract}
The Heisenberg model for spin-1 bosons in one dimension presents many different quantum phases including the famous topological Haldane phase. Here we study the robustness of such phases in front of a SU(2) symmetry breaking field as well as the emergence of unique phases. Previous studies have analyzed the effect of such uniaxial anisotropy in some restricted relevant points of the phase diagram. Here we extend those studies and present the complete phase diagram of the spin-1 chain with uniaxial anisotropy. To this aim, we employ the density matrix renormalization group (DMRG) together with analytical approaches. The complete phase diagram can be realized using ultracold spinor gases in the Mott insulator regime under a quadratic Zeeman effect.  

\end{abstract}

\maketitle

\section{Introduction}
Quite generally, ultracold atoms trapped in optical lattices provide a toolbox to reproduce various kinds of Hubbard models
\cite{Jaksch-toolbox} offering thus a unique possibility to simulate quantum magnetism~\cite{Lewenstein2007,Bloch2008,Hauke2010, Schmied2008}. 
As it is well known, Hubbard models reduce to various spin models in certain limits. Most of these limits, and even many others, are accessible with ultracold lattice atoms.

There are several ways of achieving such a situation, the simplest perhaps is the case where spin-$1/2$ is associated 
to the presence or the absence of a structureless boson in a lattice site (the, so called,
hard-core boson limit).  More generally, if one considers systems
of particles with several internal states in many body Mott
insulator states,  such systems are well described by low energy
effective Hamiltonians that incorporate super-exchange
interactions \cite{pwanderson58}, and quite generally correspond
to spin models \cite{auerbach-book,Lukin2003,
Svistunov2003,Lewenstein2007}, or composite fermions models
\cite{Svistunov2003,Lewenstein2004}. If the
internal states correspond to pseudo-spin, the resulting
Hamiltonians typically do not have special symmetries. If, on the contrary,  the
internal states correspond to the degenerate manifold of the hyperfine atomic level $F$, i.e. $\{\ket{F,m_F}\}$ of an ultracold spinor
gas, the resulting Hamiltonians, in the absence of symmetry
breaking fields, are $SU(2)$ symmetric. The corresponding spin Hamiltonians are given by powers of the
nearest neighbor Heisenberg interactions 
$H=\sum_k a_k({\bf S}_i\cdot {\bf S}_j)^k$ (for $F=1$ see for instance \cite{Imambekov}, for $F=2,3/2,5/2$ c.f. \cite{Zawitkowski}, and for
general Fermi systems c.f.  \cite{Hofstetter,congjunwureview,Hermele2009}).

The are two kinds of symmetry breaking fields that can be realized
and controlled both in atomic-ionic and in condensed matter
systems: those corresponding to the linear Zeeman effect for atoms
($z$ aligned magnetic field in condensed matter), or the quadratic
Zeeman effect (single ion anisotropy in condensed matter). The
linear Zeeman effect is important, but not so interesting, since
the magnetization is a constant of the motion, and it can
effectively be ``gauged out" \cite{Rodriguez}. Also, in experiments,
frequently magnetic fields oscillate rapidly in time, and their
effect averages to zero. This is not the case for the quadratic
Zeeman effect, which leads to more complex and interesting effects.
\cite{Stamper-Kurn,Mukerjee,Kolezhuk}
Particulary interesting are these effects in  one dimension (1D), where the role
of quantum fluctuations is enhanced.

For these reasons, considerable efforts have been devoted to the
study of the simplest case of spin-1 chains, namely the, so
called, bilinear-biquadratic spin-1 chains undergoing a quadratic
Zeeman effect. On the other hand, for the
most of the known  spin-1 alkali atomic systems, the scattering
lengths in different scattering channels are practically equal, so
that the scattering is effectively spin state independent. That
means that the systems exhibits higher symmetry,  in the case of
$F=1$ the $SU(3)$ symmetry. A similar situation  occurs with earth
alkali fermionic atoms, that can have spin $F$ and exhibit spin
state independent collisions, i.e. $SU(2F+1)$
symmetry\cite{Hermele2009}.

Very recently, Rodriguez {\it et
al.} \cite{Rodriguez} have studied the phase diagram of the bilinear-biquadratic
spin-1 chains undergoing the quadratic Zeeman effect close to the
$SU(3)$ symmetric point. These authors have used an elegant
effective field model, that allowed them to evaluate approximately
the phase boundaries for the case of vanishing magnetization in
any dimension. 
%with accuracy very well confirmed by numerical simulations. 
The study of Ref. \cite{Rodriguez} was,
however, restricted to a quite limited region of parameters. The reason is that
they consider only the region of the phase space achievable in alkaline ultracold gases by means of standard Feshbach resonances. 
 %of the
%model.  The reason for that is that the standard way of
%manipulating interactions in spinor gases explores, unfortunately,
%such a  limited interval.

The present paper complements and extends the study of Rodriguez {\it
et al.}\cite{Rodriguez} to the full range of parameters of the uniaxial spin-1 chain. 
We point out that by using the idea of employing ``quantum simulators in excited states"
\cite{Garcia-ripoll}, i.e. studying properties of the spin chain
not necessarily in the ground state, but in the state of maximal
energy, it is possible to cover the entire range of parameters for
the bilinear biquadratic spin-1 chains: from ferromagnetic
to fully anti-ferromagnetic.  The idea is here very simple: The
state of maximal energy, if gapped, is dynamically stable, and can
only undergo thermodynamic instability, which, however, requires
interactions with a reservoir etc. The maximal energy state will
thus remain stable if the system is well isolated.

The phase diagram in the spin chain is obtained numerically 
in the entire parameter region using a finite size density
matrix renormalization group (DMRG) algorithm \cite{dmrg}.  Using degenerate perturbation
theory, we show in a neat way how the emergence of the phases can be obtained from
the anisotropy of an effective XXZ spin-1/2 model.  Our results are summarized in Fig. 1, 
where for reference we mark also the region studied in~\cite{Rodriguez}. Our results agree
quantitatively with the description provided in ~\cite{Rodriguez} in this region, although we
found some significant discrepancies in the boundaries of some of the phases, as we shall
comment later.

The paper is organized as follows: in Sec. \ref{sec:model} we introduce
the model and the quantities of interest. Here we briefly review
the earlier, mainly condensed matter, literature for a non-zero anisotropy field.  
In Sec. III we initiate the study of the complete phase diagram. We discuss the results in various important limits, 
first for large negative $D$ (Sec. \ref{sec:perturbation}) where perturbation theory leads to an effective XXZ spin 1/2 model.  
The distinct emerging phases are subsequently analyzed: dimer phase (Sec. \ref{sec:dimer}); the N\'
eel phase (Sec. \ref{sec:neel});  the Haldane phase (Sec. \ref{sec:haldane}); the
ferromagnetic phases (Sec. \ref{sec:ferro}) and the critical phases (Sec.
\ref{sec:critical}).  We conclude in Sec. \ref{sec:conclusion}.

\section{The spin-1 chain in a uniaxial field: previous results}
\label{sec:model} We consider a one dimensional spin-1 chain of
length $L$ with open boundary conditions whose Hamiltonian is:
\begin{eqnarray}
\label{eq:H}
H &=&H_{BB}+H_U=
\nonumber\\
&=&J\sum_{i=1}^{L-1}\left[ \cos(\theta) \bm{S}_i\cdot\bm{S}_{i+1}+ \sin(\theta) (\bm{S}_i\cdot\bm{S}_{i+1})^2\right]
\nonumber
\\
&+&D\sum_{i=1}^L S_{zi}^2
\end{eqnarray}
where $\bm{S}_i=(S_{xi},S_{yi},S_{zi})$ are the $i$th site angular
momentum operators. The model is the sum of two terms, $H_{BB}$
and $H_U$. The former, proportional to $J$, is known as the
bilinear-biquadratic spin-1 Hamiltonian, while the latter is a
uniaxial field term proportional to $D$. In the rest of this section we review known results about this model.

\subsection{The case $D=0$}
For $D=0$, the phase
diagram of $H_{BB}$ as a function of $\theta\in[-\pi;\pi]$ is well
known (see for example \cite{Schollwock1996}). For $-\pi<\theta<-3\pi/4$ and for $\pi/2<\theta<\pi$ the
ground state is ferromagnetic with a net spontaneous magnetization
along a direction which breaks spontaneously space isotropy. For
$-3\pi/4<\theta<-\pi/4$ the ground state is characterized by a non
vanishing dimer order parameter:
\begin{equation}
\mathcal D=|H_i-H_{i+1}|
\end{equation}
 where $H_i=\cos(\theta) \bm{S}_i\cdot\bm{S}_{i+1}+ \sin(\theta) (\bm{S}_i\cdot\bm{S}_{i+1})^2$. In this phase, translational invariance is broken and singlets of neighboring spins appear. On the other hand, a nematic phase has been predicted in the region between the dimer and ferromagnetic phase \cite{nematic}. Its existence is however still debated \cite{FathSolyom1995,Buchta2005,Rizzi2005,Lauchli2006}. The difficulty in establishing  the true nature of the region close to the ferromagnetic phase is due to the fact that the dimer order parameter decays exponentially as $\theta\to-3\pi/4$.

%%%%%%%%%%%%%%%%%%%%%%%%%%%%%%%%%%%
\begin{figure}[t!]
\begin{center}
\includegraphics[scale=0.33]{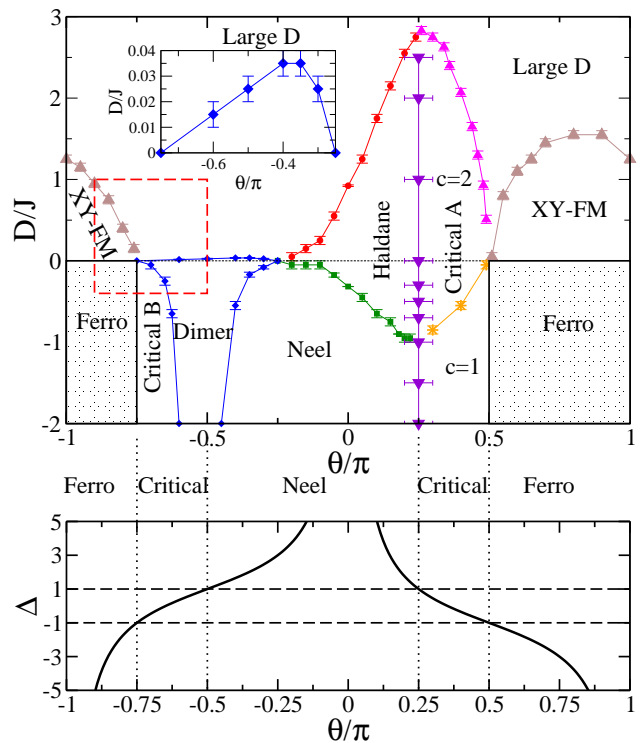}
\caption{(Color online) Top: Phase diagram of a spin-1 chain in a Zeeman quadratic field as a function of $\theta$ and $D$ obtained by means of the DMRG algorithm. The dashed red box indicates the portion of phase diagram studied in Ref.\cite{Rodriguez}. Bottom: the anisotropy $\Delta$ of the effective spin-1/2 XXZ model in the limit of large negative $D$ as a function of $\theta$ (see the results of perturbation theory  in Sec.~\ref{sec:perturbation}). The vertical dotted lines separate the resulting phases: N\'eel ($\Delta>1$), critical ($-1<\Delta<1$) and ferromagnetic ($\Delta<-1$).}
\label{fig:phase}
\end{center}
\end{figure}
%%%%%%%%%%%%%%%%%%%%%%%%%%%%%%%%%%%

For $-\pi/4<\theta<\pi/4$ the system is in the Haldane phase. This is  a topological phase, containing for $\theta=\arctan(1/3)$ the Affleck, Lieb, Kennedy, Tasaki (AKLT) point \cite{AKLT} whose ground state can be written in terms of matrix product states. The entire phase is characterized by free edge spins and a non vanishing string order parameter \cite{Rommelse, Schollwock1996}:
\begin{equation}
\label{eq:string}
O = \lim_{r\to\infty} \langle S_{zi}\exp[i\pi\sum_{j=i+1}^{i+r-1} S_{zj}]S_{zi+r}\rangle
\end{equation}
Recently, it has been shown that topological phases in one
dimensional systems can be detected by a doubly degenerate
entanglement spectrum \cite{pollmann}. In other words, the
eigenvalues of the reduced density matrix of a portion of the
chain are always doubly degenerate in a topological phase. Moreover, the
characterization of the phase transitions around the Haldane phase
in terms of the Schmidt gap, i.e. the difference of the two largest
density matrix eigenvalues, has been recently reported in
\cite{Schmidtgap}.

Finally for $\pi/4<\theta<\pi/2$ the system is in a critical,
gapless phase, characterized by zero energy modes at momentum
$q=0,\pm2\pi/3$ and has been extensively studied\cite{Lauchli2006}. The existence of these gapless modes
is reflected in a peak at $q=2\pi/3$  in the magnetic and nematic
structure factors:
\begin{equation}
S_C(q) = \frac 1L \sum_{mn} \exp[i q(m-n)]\langle C_{m}C_{n}\rangle
\end{equation}
where $C=S_z$ or $C=S_z^2$ respectively. The periodicity induced
by these momenta is also connected to the existence of trimers in
the phase. Very recently, an order parameter has been proposed
which can be experimentally observed in an optical lattice setting
\cite{dechiara-spectroscopy}.

\subsection{The case $D\neq0$}

The Hamiltonian of Eq.~\eqref{eq:H} has also been studied for
$\theta=0$ and different values of $D$
\cite{Botet,Glaus,Schulz,Papa,EspostiBoschi,Chen,Albu}. The
resulting ground state phase diagram contains three phases: in the
vicinity of $D=0$ the system is in the already mentioned Haldane
phase; for large negative values of $D$, the ground state exhibits
a spontaneous staggered magnetization:
\begin{equation}
\label{eq:staggered}
M_z^{st} =\frac 1L \sum_{n} (-1)^n S_{zn}
\end{equation}
and is therefore in an Ising-N\'eel phase;
finally, for large positive $D$ the ground state is adiabatically connected to the trivial state $\ket{00\dots 0}$ in which all the spins are in the $S_{zn}=0$ eigenstate. For this reason this phase has been dubbed the ``large $D$'' phase.

Recently, Rodriguez et al. \cite{Rodriguez} have proposed to realize
model \eqref{eq:H} for $D\neq 0$ with spinor ultracold atoms in
optical lattices. They found that by making use of Feshbach
resonances in ${}^{87}$Rb atoms, the interval
$\theta\in(-\pi+\arctan\frac 13;-\pi/2)$ around the SU(3) point can be realized. They  
studied the corresponding phase diagram in 1D, 2D and 3D, finding unique phases: an XY-ferromagnetic phase for $D>0$ and $\theta<-3\pi/4$ and an XY-nematic phase for $D<0$ and $\theta\in [-3\pi/4;-\pi/2]$.

\section{The complete phase diagram}
In this paper we study the complete phase diagram of Hamiltonian~\eqref{eq:H} in the interval $\theta\in[-\pi;\pi]$ and $D/J\in[-2;3]$. 
Our main result is summarized in Fig.~\ref{fig:phase} showing the resulting phase diagram.
 In the limit of large negative $D/J$ we apply perturbation theory to reconstruct the phase diagram. In the general case, we conduct extensive numerical simulations using the finite size density matrix
renormalization group (DMRG) algorithm (see for example
\cite{dmrg} and references therein) with open boundary conditions. The number of block states retained has been always chosen so that
the discarded weight, connected to the algorithm accuracy, is less
than $10^{-6}$ or until convergence is reached. Typically we used between $100$ and $250$ states.
  In the remainder of the paper we discuss  the different phases individually
and the transitions separating them.

\subsection{Limit of large negative $D$}
\label{sec:perturbation}
Let us start by considering the limit of large negative $D$ so that we can apply perturbation theory assuming $|J/D|\ll 1$. For $J=0$ and negative $D$ the ground state of $H_U$ is degenerate and pertains to the manifold spanned by the states:
\begin{equation}
 \ket{\epsilon_1\epsilon_2\dots\epsilon_L}
\end{equation}
where $\epsilon_i=\pm 1$. The unperturbed ground state energy is simply $DL<0$ where $L$ is the size of the chain. In this limit, $H_{BB}$ acts as a perturbation and we can apply first order degenerate perturbation theory. This consists in diagonalizing the perturbation Hamiltonian in the degenerate subspace.
Projecting the Hamiltonian $H_{BB}$ on this manifold leads to the effective spin-$1/2$ Hamiltonian:
\begin{eqnarray}
\tilde H_{BB} &=& \frac 32 J L \sin\theta +J\sum_i \left[\frac 12 \sin\theta (\sigma^x_{i}\sigma^x_{i+1}+\sigma^y_{i}\sigma^y_{i+1})+
\right .
\nonumber\\
&+&\left .\frac 12(2\cos\theta-\sin\theta)\sigma^z_{i}\sigma^z_{i+1}\right]=E_0(\theta)+H_{XXZ}
\end{eqnarray}
where $\sigma^x_{i},\sigma^y_{i}$ and $\sigma^z_{i}$ are the spin-$1/2$ Pauli matrices for the $i$th spin in the basis $\{\ket{1},\ket{-1}\}$. 

Therefore, apart from a constant energy shift $E_0(\theta)$, for large negative $D$ the system is described by an effective $XXZ$ spin-$1/2$ model with anisotropy $\Delta=(2\cos\theta-\sin\theta)/|\sin\theta|$. As shown in Fig.~\ref{fig:phase}, the quantity $\Delta(\theta)$ as a function of $\theta$ spans twice the real axis when $\theta$ changes from $-\pi$ to $\pi$.
The properties of the $XXZ$ models are well known \cite{Schollwock_book}: it is ferromagnetic for $\Delta<-1$, it is in the critical $XY$ phase for $-1<\Delta<1$ and in the Ising-N\'eel phase for $\Delta>1$.
We can, therefore, reconstruct the phase diagram in this regime as a function of $\theta$ as sketched in  Fig.~\ref{fig:phase}. The system is in the N\'eel phase for $-\pi/2<\theta<\pi/4$, in the ferromagnetic phase for $-\pi<\theta<-3\pi/4$ and for $\pi/4<\theta<\pi$ and in the critical $XY$ phase for $-3\pi/4<\theta<-\pi/2$ and for $\pi/4<\theta<\pi/2$.
We use these analytical results as a guide for the numerical simulations in the limit of  large negative $D$.
Notice that in the opposite limit of large positive $D$, the ground state is non degenerate and given by the product state in which all the spins are in the state $\ket{0}$. In this case, perturbation theory gives poor results \cite{Papa}. In the following sections, we analyze separately each phase characterizing it with appropriate orders.

\subsection{The dimer phase}
\label{sec:dimer}
We begin our numerical analysis with the dimer phase. As said in Sec.~\ref{sec:model} this phase is characterized by the presence of the dimer order parameter $\mathcal D$. In order to obtain results in the thermodynamic limit we computed $\mathcal D$, using DMRG,  for different system sizes $L=96,132,168, 204$ and then performed a finite size extrapolation with a quadratic function:
\begin{equation}
\label{eq:extrapolation}
\mathcal D\left(\frac 1L\right) = \mathcal D_\infty +\frac {b_1}{L}+ \frac {b_2}{L^2}
\end{equation}
%%%%%%%%%%%%%%%%%%%%%%%%%%%%%%%%%%%
\begin{figure}[t!]
\includegraphics[scale=0.47]{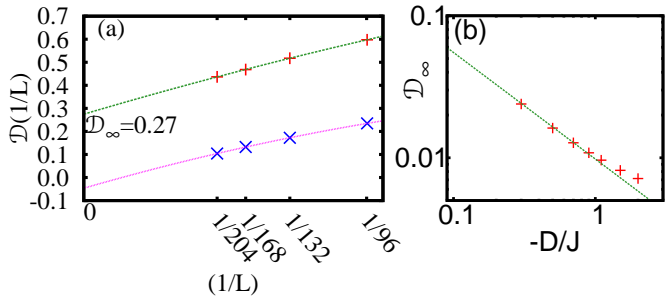}
\caption{(Color online) (a) Finite size extrapolation of the dimer order paramter for $\theta=-0.4\pi$ and two values of $D$: $D=-0.2J$ (plus) for which the extrapolated result is $\mathcal D_\infty=0.27$ and $D=-0.7J$ (crosses) for which the extrapolated value is negative and therefore $\mathcal D_\infty=0$. Also shown are the corresponding fitting functions. (b) The extrapolated dimer order parameter for $\theta=-0.6\pi$ and large negative $D$. The plot is shown as a function of $-D/J$ and in a log-log plot. The data are compared with a power-law with exponent $0.75$ (solid line).}
\label{fig:dimer}
\end{figure}
%%%%%%%%%%%%%%%%%%%%%%%%%%%%%%%%%%%

Two examples of such extrapolation are shown in Fig.~\ref{fig:dimer}(a) for $\theta=-0.4\pi$: (i) for $D=-0.2J$ the extrapolated dimer order is $\mathcal D_\infty=0.27 J$ and therefore we conclude that this point is in the dimer phase; (ii) for $D=-0.7J$ the extrapolated result is zero and therefore the point is not in the dimer phase. 
We have compared our extrapolated result in the point $D=0$ and $\theta=-\pi/2$ with the analytical result valid in the thermodynamic limit $\mathcal D=1.1243$ found in Ref.\cite{Xian} (see also \cite{FathSolyom1995}). Our result is $\mathcal D_\infty = 1.13$ and the less than $1\%$ relative error gives an idea of the quality of our numerical simulations. 

With this technique we reconstruct the dimer phase as shown in Fig.~\ref{fig:phase}. Surprisingly enough, the dimer phase persists even at $D=-2J$.  Although the dimer order parameter is different from zero, it is really small as shown in Fig.~\ref{fig:dimer}(b) and decays not faster than algebraically to zero as $D\to-\infty$. This is consistent with our predictions from perturbation theory that dimerization is zero in this limit. We cannot however exclude that the border of the dimer phase occurs for a finite value of $D/J<-2$. This result does not agree with Ref.~\cite{Rodriguez} which for $\theta=-0.5\pi$ predicts the transition around $D=-0.3 J$.

We analyze the robustness of the dimer phase for positive $D$ (see the inset of Fig.~\ref{fig:phase}): for example for $\theta=-0.5\pi$ the dimer order parameter is already zero for $D\geq 0.05 J$. Again, this is in clear contrast with the results of Ref.~\cite{Rodriguez} which predicts the transition for $D\sim 0.6 J$ and a cusp-like behavior which we do not observe. The disagreement could be due to the small system sizes taken in Ref.~\cite{Rodriguez}. The finite size effects are in fact evident in Fig.~\ref{fig:dimer}(a), showing that a careful extrapolation to the thermodynamic limit is in order.

Notice that the procedure we use here is reliable only away from criticality and serves us to approximately locate the phase boundaries. To get more accurate results for the critical points as well as critical exponents we will make use in Sec.~\ref{sec:neel} of finite size scaling theory.

We leave as an open question the behavior of the dimer phase boundaries close to the point $D=0$ and $\theta=-3/4\pi$ where the gap and the dimer order parameter are expected to decrease exponentially. Accurate boundaries are therefore hard to get with our code and probably more sophisticated tensor networks techniques are required.

\subsection{The N\'eel phase}
\label{sec:neel}
As found in previous works\cite{Botet,Glaus,Schulz,Papa,EspostiBoschi,Chen,Albu} and predicted by our perturbation theory results, in the asymptotic limit $D\to-\infty$, a N\'eel phase occurs. This phase is characterized by a non zero spontaneous staggered magnetization defined in Eq.~\eqref{eq:staggered} which we use as the order parameter.
Similar to the dimer phase, to locate the boundaries of the N\'eel phase, we extrapolate  the staggered magnetization to the thermodynamic limit. 
In order to compare our results to precise quantum Monte Carlo simulations\cite{Albu} we make use of finite size scaling (FSS) theory \cite{FSS} to obtain the corresponding critical exponents.
In the so called scaling regime, the dependence of an observable $Q$, when a parameter $g$ of the Hamiltonian (in our case $\theta$ or $D$) varies close to the critical point $g_c$, is given by:
\begin{equation}
Q\simeq L^{-\beta_Q/\nu} f_Q\left(|g-g_c| L^{1/\nu} \right)
\end{equation}
where $\nu$ characterizes the divergence of the correlation length when approaching the critical point: $\xi\sim|g-g_c|^{-\nu}$ and $\beta_Q$ is the order parameter exponent: $Q\sim |g-g_c|^{\beta_Q}$ in the ordered phase.
To locate the critical point $g_c$ we plot $Q L^{\alpha_1}$ for different sizes $L$ as a function of $g$ and change $\alpha_1$ until there is a crossing of all the curves. This procedure will give us $g_c$ at the crossing point and $\beta_Q/\nu=\alpha_1$. In order to compute $\beta_Q$, we plot $Q L^{\beta_Q/\nu}$ as a function of $(g-g_c)L^{\alpha_2}$ and change $\alpha_2$ until we observe collapse of the data for all lengths $L$. In this way we extract $\beta=\alpha_1/\alpha_2$ and $\nu=1/\alpha_2$.

Care must be taken, because in a finite size system the $Z_2$ symmetry of the Hamiltonian is not spontaneously broken. To circumvent this technical issue, we add a small magnetic field acting only on the first spin, corresponding to an additional term  $0.01JS_{z1}$.

\begin{figure}[t!]
\begin{center}
\includegraphics[scale=0.7]{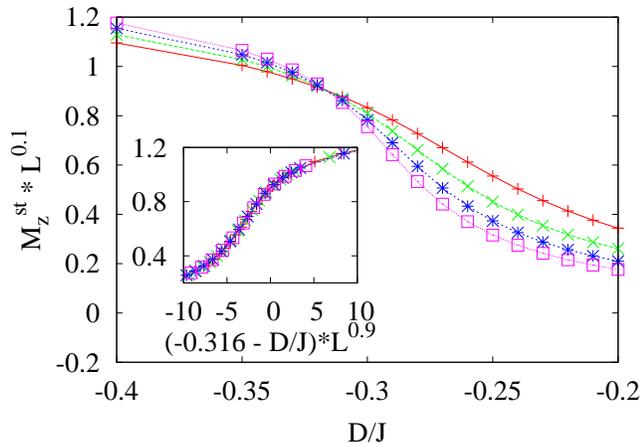}
\caption{(Color online) Finite size scaling of the staggered magnetization $M_z$ for lengths: $L=96,132,168,204$ for $\theta=0$ close to the critical point separating the N\'eel and Haldane phases. Inset: the corresponding collapse plot (see text).}
\label{fig:theta0}
\end{center}
\end{figure}

We use this method for $\theta=0$ so that we can compare our results with Ref.~\cite{Albu}, which employs quantum Monte Carlo simulations and in which the estimate for the critical point is at $D_c^{QM}=-0.316 J$ and for the exponent for the staggered magnetization $\beta_{M_z}^{QM}=0.147$ while  $\nu^{QM}= 1.01$ which are not far from the universal exponents of the 2D classical Ising model  ($\beta_{M_z}=1/8$ and  $\nu= 1$). Our results for the staggered magnetization are shown in Fig.~\ref{fig:theta0}. Crossing and collapse of the data for lengths $L=96,132,168,204$ happen for $D_c=-0.316 J$, $\beta_{M_z}=0.11\pm0.01$, $\nu= 1.1\pm0.1$ which agree with Ref.~\cite{Albu} and with the Ising universality class. We checked that for other fixed values of $\theta$, the critical points, as $D$ changes, are also of the Ising type.
%%%%%%%%%%%%%%%%%%%%%%%%%%%%%%%%%%%
%%%%%%%%%%%%%%%%%%%%%%%%%%%%%%%%%%%

\subsection{The Haldane phase}
\label{sec:haldane}
The Haldane phase is the archetypal example of a topological phase in one dimensional spin systems.
In this paper we characterize it using two key quantities: the string order parameter defined in Eq.~\eqref{eq:string} and the degeneracy of the entanglement spectrum defined as follows. After finding the ground state $\ket{\psi_G}$ of the system using DMRG, we construct the reduced density matrix of half chain:
\begin{equation}
\rho_L=\textrm{Tr}_R \ket{\psi_G}\bra{\psi_G}
\end{equation}
The entanglement spectrum is
given by the set of eigenvalues $\lambda_i$ 
of the density matrix $\rho_L$  such that $\sum_i\lambda_i=1$ and ordered in decreasing order.
A topological phase, such as Haldane,  is signaled by a doubly degenerate spectrum. A handful quantity for detecting the Haldane phase is the Schmidt gap, given by the difference of the two largest eigenvalues of $\rho_L$:
$
\Delta\lambda = \lambda_1-\lambda_2$. 

\begin{figure}[t!]
\begin{center}
\includegraphics[scale=0.33]{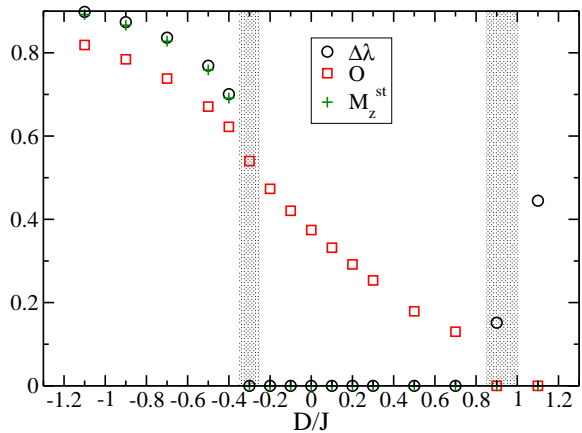}
\caption{(Color online) The Schmidt gap, the string order parameter and the staggered magnetization for $\theta=0$ are shown as a function of $D/J$. The data have been obtained from a quadratic extrapolation for the lengths $L=96,132,168,204$ (see Eq.~\eqref{eq:extrapolation}). The shaded area represents the scaling regions close to the critical points where the quadratic extrapolation does not work.}
\label{fig:string}
\end{center}
\end{figure}

We studied the robustness of the topological phase in presence of the SU(2) Zeeman quadratic field using both the Schmidt gap and the string order parameter.
In Fig.~\ref{fig:string}, we display our results for $\theta=0$ and for positive and negative values of $D/J$. Both the Schmidt gap and the string order $O$ clearly identify the Haldane phase, however one has to be careful since $O$ is different from zero also in the N\'eel phase. This is evident if one considers the perfect N\'eel state $\ket{+1,-1,+1,-1\dots}$. This ambiguity is resolved by looking at the staggered magnetization $M_z^{st}$ also reported in the plot. Contrary to the N\'eel phase, the string order goes instead to zero in the large D phase, and at the same time the Schmidt gap reopens as expected.
Fig.~\ref{fig:string} shows that the Schmidt gap and the staggered magnetization scale very similarly. In fact in a recent work \cite{Schmidtgap} we have shown that, in proximity of the N\'eel-Haldane boundary, the two quantities $M_z^{st}$ and $\Delta\lambda$ scale with the same critical exponents and can be therefore considered to be two possible order parameters for the description of this transition.

Finally, we have compared our numerical results approaching the point $\theta=\pi/4$ and $D=0$  known as the Uimin-Lai-Sutherland (ULS) model \cite{ULS} with exact analytical calculations. The ULS model in the presence of a linear and a quadratic field was studied in \cite{Schmitt} and the two critical values of $D$ are $D_+=2.8284 J$ and $D_-=-0.9800 J$. Notice that these values have been obtained by dividing the corresponding results in \cite{Schmitt} by a factor $\sqrt 2$ because of the different Hamiltonian definition in our model.  Our estimates are $D_+=2.83\pm0.05$ and $D_-=-0.95\pm0.05$ which are in excellent agreement with the analytical predictions.

%%%%%%%%%%%%%%%%%%%%%%%%%%%%%%%%%%%
%%%%%%%%%%%%%%%%%%%%%%%%%%%%%%%%%%%

\subsection{The Ferromagnetic phases}
\label{sec:ferro}
%{\it We should discuss this part of the ferromagnetic phase, since the boundaries of the ferromagnetic phase are supposed to be first order phase transitions and given simply by the lines: $D=0$, $\theta=-3\pi/4$ and $\theta=\pi/4$}.

The system is in a trivial ferromagnetic phase for $D<0$ and for $-\pi<\theta<-3\pi/4$ and $\pi/2<\theta<\pi$. In this Ising ferromagnetic phase, the model is characterized by a non zero magnetization $M_z= 1/L \sum_{n} S_{zn}$ along $z$ (except along the line $D=0$, where the magnetization could be in any direction).

\begin{figure}[t!]
\begin{center}
\includegraphics[scale=0.33]{5a}
\includegraphics[scale=0.7]{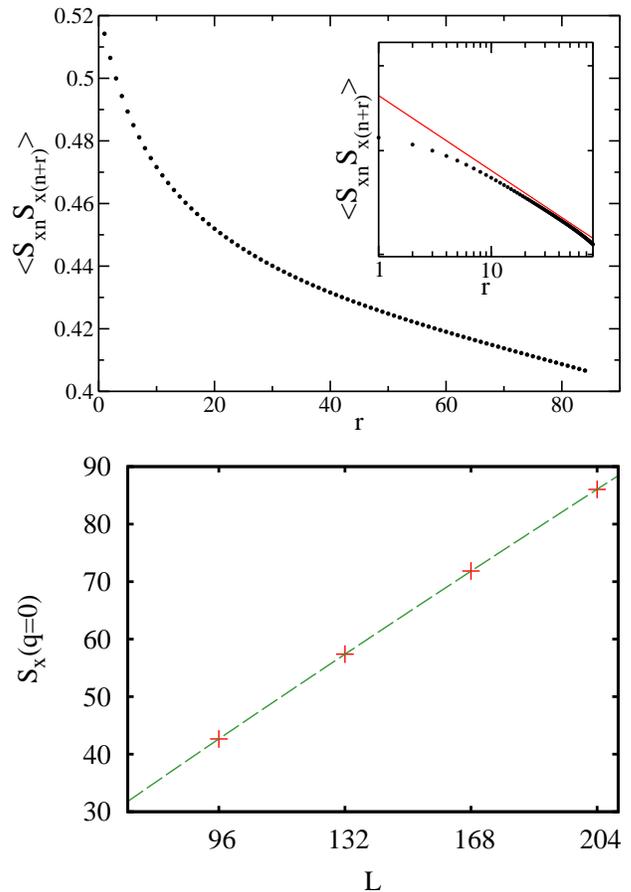}
\caption{(Color online) Top: In-plane correlations $\langle S_{xn}S_{x(n+r)}\rangle$ as a function of the spin distance $r$ for $L=168,\theta=-0.8\pi, D/J=0.1$. Notice that $r$ is limited to the interval between $1$ and $L/2$ to avoid boundary effects. Inset the same plot in logarithmic scale. The data are compared with a powerlaw $\sim 1/r^{0.07}$. Bottom: the magnetic structure factor for the $XX$ correlations $S_x (q=0)$ for zero momentum as a function of the system size $L$. The solid line as been obtained with a power-law fit $\sim L^{(1-\alpha)}$ and the best-fit value is $\alpha= 0.07$.}
\label{fig:xy}
\end{center}
\end{figure}

On the contrary, for the same values of $\theta$ but for $D>0$, the net magnetization is zero along any direction. However for small values of $D$ the system is in an $XY$-Ferromagnetic state as predicted in Ref.~\cite{Rodriguez}. Although the ``in plane'' magnetization is zero, since the system is one dimensional, the correlations in the $XY$ plane decay algebraically as shown in Fig.~\ref{fig:xy}. To find the boundary between the XY-Ferromagnetic phase and the large $D$ phase we study the magnetic structure factor for the $XX$ correlations:
\begin{equation}
S_x(q) = \frac 1L \sum_{mn} \exp[i q(m-n)]\langle S_{xm}S_{xn}\rangle
\end{equation}
In the $XY$-Ferromagnetic phase, the correlations are expected to decay as $\langle S_{xn}S_{x(n+r)}\rangle\sim r^{-\alpha}$.
It is very hard to extract an accurate value of $\alpha$ by fitting the correlation functions. Instead, by considering that the magnetic structure factor diverges for $q=0$ with the system size:  
 $S_x(q=0)\sim L^{1-\alpha}$, one can better estimate $\alpha$. In the large $D$ phase, correlations decay exponentially with a finite correlation length and the corresponding magnetic structure factor  $S_x(q=0)$ is independent of $L$.  Therefore, in order to distinguish the $XY$-Ferromagnetic phase from the large $D$ phase, it is sufficient to study the behavior of $S_x(q=0)$ with $L$ and check whether it diverges or not. An example of such analysis is shown in Fig.~\ref{fig:xy} for $\theta=-0.8\pi$ and $D/J=0.1$. In the top panel the correlations $\langle S_{xn}S_{x(n+r)}\rangle$ are shown for $n=L/4$ as a function of $r$. The value of $n$ has been chosen so as to minimize finite size effects from the boundaries. In the bottom panel, we show the magnetic structure factor at zero momentum as a function of the system size. Using a power-law fit we find $\alpha=0.07$.

%%%%%%%%%%%%%%%%%%%%%%%%%%%%%%%%%%%
%%%%%%%%%%%%%%%%%%%%%%%%%%%%%%%%%%%

\subsection{The critical phases}
\label{sec:critical}
\begin{figure}[t!]
\begin{center}
\includegraphics[scale=0.6]{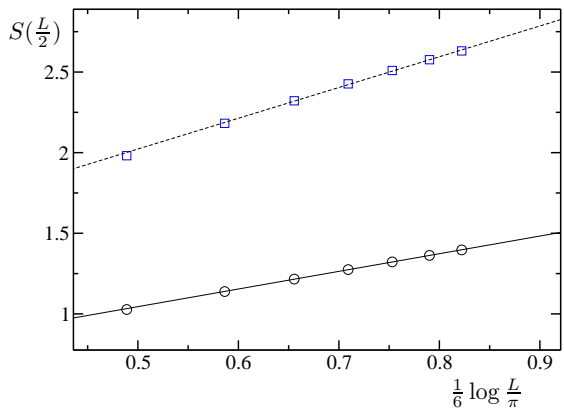}
\caption{(Color online) von Neumann  entropy at half chain for $\theta=0.3\pi$ and two values of $D$: $D=-2 J$ (circles) and $D=0$ (squares). The entropy is plotted as a function of $\frac 16 \log\frac L\pi$ so that the entropy should behave as a straight line whose slope is $c$. The solid and dashed lines are the corresponding linear fits with slope $1.09$ and $1.91$ respectively.}
\label{fig:entropy}
\end{center}
\end{figure}

The phase diagram contains also two interesting critical phases as shown in Fig.~\ref{fig:phase} denoted as Critical A and Critical B.
We find interestingly enough that Critical A has at least two different central charges $c\simeq 1$ and $c\simeq 2$, each identifying a different universality class \cite{CardyBook}. We base our results on the scaling of the von Neumann entropy with the size of a partition of the chain. 
%For $\pi/4<\theta<\pi/2$ this critical gapless phase extends from negative to positive $D$. For $D=0$ the phase is well known and characterized by soft mode at momenta $k=\pm 2\pi/3$ which result in period-$3$ oscillations in the correlation functions \cite{Lauchli2006,dechiara-spectroscopy} as we mentioned in the Sec.~\ref{sec:model}. 
 If $\rho_l$ is the reduced density matrix of a block of spins of length $l$ when the global state of $L$ spins is the ground state, then the von Neumann entropy is given by:
\begin{equation}
S(l)=-\textrm{Tr}\rho_l\log\rho_l
\end{equation}
For critical points in one dimensional quantum spin chains, the von Neumann entropy scales logarithmically with $l$, Ref.~\cite{Calabrese04}:
\begin{equation}
\label{eq:esel}
S(l)=\frac c6 \log\left[\frac L\pi \sin(\frac{\pi l}{L})\right] + A
\end{equation}
where $c$ is the central charge of the corresponding conformal field theory and $A$ is a non universal constant.
Our results for $S(L/2)$ computed at half chain are shown in Fig.~\ref{fig:entropy} 
 for $D=-2 J$ and for $D=0$ and $\theta=0.3\pi$. The best fits of function \eqref{eq:esel} for $D=-2 J$ and for $D=0$ give $c=1.09$ and $c=1.91$ respectively for the central charge. 

Moreover, our results are compatible with previous results\cite{Aguado09} which obtained at $\theta=\pi/4$ and $D=0$ the central charge $c=1.995$, which is compatible with the expected result $c=2$.
On the other hand we expect, in the limit $|D/J|\gg 1$ and $D$ negative, the central charge to become $c=1$, in agreement with our numerical results. In fact, our perturbative analysis of Sec.~\ref{sec:perturbation} predicts the model to be described by an effective $XXZ$ spin $1/2$ chain in the critical region. A numerical estimate for the central charge for this model is very close to $c=1$ as shown in Ref.~\cite{DeChiara06}.
By studying the change in the estimated central charge for fixed $\theta$ and as a function of $D$ we find the line separating the two regions. Whether a crossover or a phase transition separate the two regions, remains an open problem.

%%%%%%%%%%%%%%%%%%%%%%%%%%%%%%%%%%%
%%%%%%%%%%%%%%%%%%%%%%%%%%%%%%%%%%%
  
The boundaries of the  Critical B  phase are found by the disappearance of the dimer order parameter and the absence of magnetization. The nematic order parameter $Q=\langle S_{zi}^2\rangle-2/3$ is different from zero in this phase. However, this is not sufficient to signal unamibiguously this phase since $Q$ is different from zero also in the dimer phase for $D<0$. It is expected, as explained in Ref.~\cite{Rodriguez}, that this phase is characterized by enhanced nematic correlations. This should be also true for the Critical A phase in the limit of large negative $D$. Since these phases are gapless, accurate DMRG calculations are computationally hard to perform. Nevertheless, we compute the block entropy in the Critical B phase for $\theta=-0.7\pi$ and $D/J=-0.5$ and obtain a central charge estimate: $c=0.99$ which is compatible with the predicted result $c=1$ valid in the limit of large negative $D$. This result is in agreement with Ref.~\cite{Rodriguez} in which a Berezinskii-Kosterlitz-Thouless transition is predicted (see also \cite{Chen}).

\section{Summary}
\label{sec:conclusion}
Summarizing, we have studied the complete phase diagram of the spin-$1$ chain in the presence of a uniaxial symmetry breaking field. Our study is
based on the density matrix renomalization group (DMRG) algorithm where finite size effects have been carefully taked into account to determine the
phase boundaries with accuracy. Our results compare well to previous results of some points in the phase diagram calculated by quantum Monte Carlo
algorithms, and agree qualitatively with the recent study presented in \cite{Rodriguez}. Furthermore, our numerical results for large negative 
value of the anisotropy, coincide with those predicted by an effective spin-1/2 XXZ model that we obtain as a limit of degenerated perturbation theory.
Finally, we remark that the whole phase diagram is achievable with ultracold spinor gases by realizing that ground states of antiferromagnetic 
phases can be mapped to the maximal energy states of ferromagnetic phases \cite{Garcia-ripoll}.

{\it Acknowledgments.--} We thank R. Fazio, A. Muramatsu and T. Vekua for enlightening discussions. We acknowledge
support from the Spanish MICINN (Juan de la Cierva,
FIS2008-01236, FIS2008-00784 and QOIT-Consolider
Ingenio 2010), Generalitat de Catalunya Grant No.
2005SGR-00343, ERC (QUAGATUA), EU (AQUTE,
NAMEQUAM), NSF (PHY005-51164). We used the DMRG code available at \url{http://www.dmrg.it}.
We acknowledge the Kavli Institute of Theoretical Physics, Santa Barbara, 
California, USA, where part of this work has been performed.

\end{document}